\PassOptionsToPackage{unicode}{hyperref}
\PassOptionsToPackage{hyphens}{url}
\PassOptionsToPackage{dvipsnames,svgnames,x11names}{xcolor}
\documentclass[
  letterpaper,
  DIV=11,
  numbers=noendperiod]{scrartcl}

\usepackage{amsmath,amssymb}
\usepackage{lmodern}
\usepackage{iftex}
\ifPDFTeX
  \usepackage[T1]{fontenc}
  \usepackage[utf8]{inputenc}
  \usepackage{textcomp} 
\else 
  \usepackage{unicode-math}
  \defaultfontfeatures{Scale=MatchLowercase}
  \defaultfontfeatures[\rmfamily]{Ligatures=TeX,Scale=1}
\fi
\IfFileExists{upquote.sty}{\usepackage{upquote}}{}
\IfFileExists{microtype.sty}{
  \usepackage[]{microtype}
  \UseMicrotypeSet[protrusion]{basicmath} 
}{}
\makeatletter
\@ifundefined{KOMAClassName}{
  \IfFileExists{parskip.sty}{%
    \usepackage{parskip}
  }{
    \setlength{\parindent}{0pt}
    \setlength{\parskip}{6pt plus 2pt minus 1pt}}
}{
  \KOMAoptions{parskip=half}}
\makeatother
\usepackage{xcolor}
\ifx\paragraph\undefined\else
  \let\oldparagraph\paragraph
  \renewcommand{\paragraph}[1]{\oldparagraph{#1}\mbox{}}
\fi
\ifx\subparagraph\undefined\else
  \let\oldsubparagraph\subparagraph
  \renewcommand{\subparagraph}[1]{\oldsubparagraph{#1}\mbox{}}
\fi

\providecommand{\tightlist}{%
  \setlength{\itemsep}{0pt}\setlength{\parskip}{0pt}}\usepackage{longtable,booktabs,array}
\usepackage{calc} 
\usepackage{etoolbox}
\makeatletter
\patchcmd\longtable{\par}{\if@noskipsec\mbox{}\fi\par}{}{}
\makeatother
\IfFileExists{footnotehyper.sty}{\usepackage{footnotehyper}}{\usepackage{footnote}}
\makesavenoteenv{longtable}
\usepackage{graphicx}
\makeatletter
\def\maxwidth{\ifdim\Gin@nat@width>\linewidth\linewidth\else\Gin@nat@width\fi}
\def\maxheight{\ifdim\Gin@nat@height>\textheight\textheight\else\Gin@nat@height\fi}
\makeatother
\setkeys{Gin}{width=\maxwidth,height=\maxheight,keepaspectratio}
\makeatletter
\def\fps@figure{htbp}
\makeatother
\newlength{\cslhangindent}
\newlength{\csllabelwidth}
\newlength{\cslentryspacingunit} 
\newenvironment{CSLReferences}[2] 
 {
  \setlength{\parindent}{0pt}
  \ifodd #1
  \let\oldpar\par
  \def\par{\hangindent=\cslhangindent\oldpar}
  \fi
  \setlength{\parskip}{#2\cslentryspacingunit}
 }%
 {}
\usepackage{calc}

\newcommand{\CSLLeftMargin}[1]{\parbox[t]{\csllabelwidth}{#1}}
\newcommand{\CSLRightInline}[1]{\parbox[t]{\linewidth - \csllabelwidth}{#1}\break}

\KOMAoption{captions}{tableheading}
\makeatletter
\@ifpackageloaded{caption}{}{\usepackage{caption}}
\AtBeginDocument{%
\ifdefined\contentsname
  \renewcommand*\contentsname{Table of contents}
\else
  \newcommand\contentsname{Table of contents}
\fi
\ifdefined\listfigurename
  \renewcommand*\listfigurename{List of Figures}
\else
  \newcommand\listfigurename{List of Figures}
\fi
\ifdefined\listtablename
  \renewcommand*\listtablename{List of Tables}
\else
  \newcommand\listtablename{List of Tables}
\fi
\ifdefined\figurename
  \renewcommand*\figurename{Figure}
\else
  \newcommand\figurename{Figure}
\fi
\ifdefined\tablename
  \renewcommand*\tablename{Table}
\else
  \newcommand\tablename{Table}
\fi
}
\@ifpackageloaded{float}{}{\usepackage{float}}
\floatstyle{ruled}
\@ifundefined{c@chapter}{\newfloat{codelisting}{h}{lop}}{\newfloat{codelisting}{h}{lop}[chapter]}
\floatname{codelisting}{Listing}

\makeatother
\makeatletter
\@ifpackageloaded{caption}{}{\usepackage{caption}}
\@ifpackageloaded{subcaption}{}{\usepackage{subcaption}}
\makeatother
\makeatletter
\@ifpackageloaded{tcolorbox}{}{\usepackage[many]{tcolorbox}}
\makeatother
\makeatletter
\@ifundefined{shadecolor}{\definecolor{shadecolor}{rgb}{.97, .97, .97}}
\makeatother
\ifLuaTeX
  \usepackage{selnolig}  
\fi
\IfFileExists{bookmark.sty}{\usepackage{bookmark}}{\usepackage{hyperref}}
\IfFileExists{xurl.sty}{\usepackage{xurl}}{} 
\hypersetup{
  pdftitle={Unification of species, gene, and cell trees for single-cell expression analyses},
  colorlinks=true,
  linkcolor={blue},
  filecolor={Maroon},
  citecolor={Blue},
  urlcolor={Blue},
  pdfcreator={LaTeX via pandoc}}

\title{Unification of species, gene, and cell trees for single-cell
expression analyses}
\author{}
\date{}

\begin{document}
\maketitle
\ifdefined\Shaded\renewenvironment{Shaded}{\begin{tcolorbox}[sharp corners, breakable, frame hidden, interior hidden, borderline west={3pt}{0pt}{shadecolor}, boxrule=0pt, enhanced]}{\end{tcolorbox}}\fi

Samuel H. Church, Jasmine L. Mah, Casey W. Dunn

Department of Ecology and Evolutionary Biology, Yale University, New
Haven, CT, 06511

\hypertarget{abstract}{%
\section{Abstract}\label{abstract}}

Comparisons of single-cell RNA sequencing (scRNA-seq) data across
species can reveal links between cellular gene expression and the
evolution of cell functions, features, and phenotypes. These comparisons
invoke evolutionary histories, as depicted with phylogenetic trees, that
define relationships between species, genes, and cells. Here we
illustrate a tree-based framework for comparing scRNA-seq data, and
contrast this framework with existing methods. We describe how we can
use trees to identify homologous and comparable groups of genes and
cells, based on their predicted relationship to genes and cells present
in the common ancestor. We advocate for mapping data to branches of
phylogenetic trees to test hypotheses about the evolution of cellular
gene expression. We describe the kinds of data that can be compared, and
the types of questions that each comparison has the potential to
address. Finally, we reconcile species phylogenies, gene phylogenies,
cell phylogenies, and cell lineages as different representations of the
same concept: the tree of cellular life. By integrating phylogenetic
approaches into scRNA-seq analyses, we can overcome challenges for
building informed comparisons across species, and robustly test
hypotheses about gene and cell evolution.

\hypertarget{introduction}{%
\section{Introduction}\label{introduction}}

Single-cell RNA sequencing (scRNA-seq) generates high-dimensional gene
expression data from thousands of cells from an organ, tissue, or
body\textsuperscript{1}. Single-cell expression data are increasingly
common, with new animal cell atlases released every
year\textsuperscript{2--6}. The next step is to compare atlases across
species\textsuperscript{2}, to identify the dimensions in which these
results differ, and to associate these differences with other features
of interest\textsuperscript{7}. Because all cross-species comparisons
are inherently evolutionary comparisons, there is an opportunity to
integrate approaches from the field of evolutionary biology, and
especially phylogenetic biology\textsuperscript{8}. Drawing concepts,
models, and methods from these fields will overcome central challenges
with comparative scRNA-seq, especially in how to draw coherent
comparisons over thousands of genes and cells across species. In
addition, this synthesis will help avoid the unnecessary reinvention of
analytical methods that have already been rigorously tested in
evolutionary biology for other types of data, such as morphological and
molecular data.

Comparative gene expression has been used for decades to answer
evolutionary questions (e.g.~\emph{How are changes in gene expression
associated with the evolution of novel functions and
phenotypes?})\textsuperscript{9}. Single-cell RNA sequencing represents
a massive increase in the scale of these experiments\textsuperscript{1},
from working with a few genes or a few tissues, to assays of the entire
transcriptome, across thousands of cells in a dissociation experiment.
Comparative scRNA-seq therefore allows us to scale up our evolutionary
questions, for example:

\begin{itemize}
\tightlist
\item
  \emph{How has the genetic basis of differentiation evolved across cell
  populations and over time?}
\item
  \emph{What kinds of cells and gene expression patterns were likely
  present in the most recent common ancestor?}
\item
  \emph{What changes in cell transcriptomes are associated with the
  evolution of new ecologies, life-histories, or other features?}
\item
  \emph{How much variation in cellular gene expression do we observe
  over evolutionary time? Which changes in gene expression are
  significant (i.e.~larger or smaller than we expect by chance)?}
\item
  \emph{Which genes show patterns of correlated expression evolution?
  Can evolutionary screens detect novel interactions between genes?}
\end{itemize}

For these comparisons, we seek to compare and analyze the results of
individual scRNA-seq experiments across species. These experiments
generate matrices of count data, with measurements along two axes:
cells, and genes (Figure 1). Comparative scRNA-seq adds a third axis:
species. At first glance, we might seek to align scRNA-seq matrices
across species, creating a three-dimensional tensor of cellular gene
expression. But neither genes nor cells are expected to share a
one-to-one correspondence across species. In the case of genes, gene
duplication (leading to paralogous relationships) and gene loss are
rampant\textsuperscript{10}. In the case of cells, there is rarely
justification for equating two individual cells across species, instead,
we typically consider similarity across populations of cells (``cell
types'')\textsuperscript{11}. Therefore to align matrices, we must first
find the appropriate system of grouping these dimensions. This is
essentially a question of homology\textsuperscript{12}: which genes and
cell types are homologous, based on their relationship to predicted
genes and cell types in the common ancestor.

\begin{figure}

{\centering \includegraphics{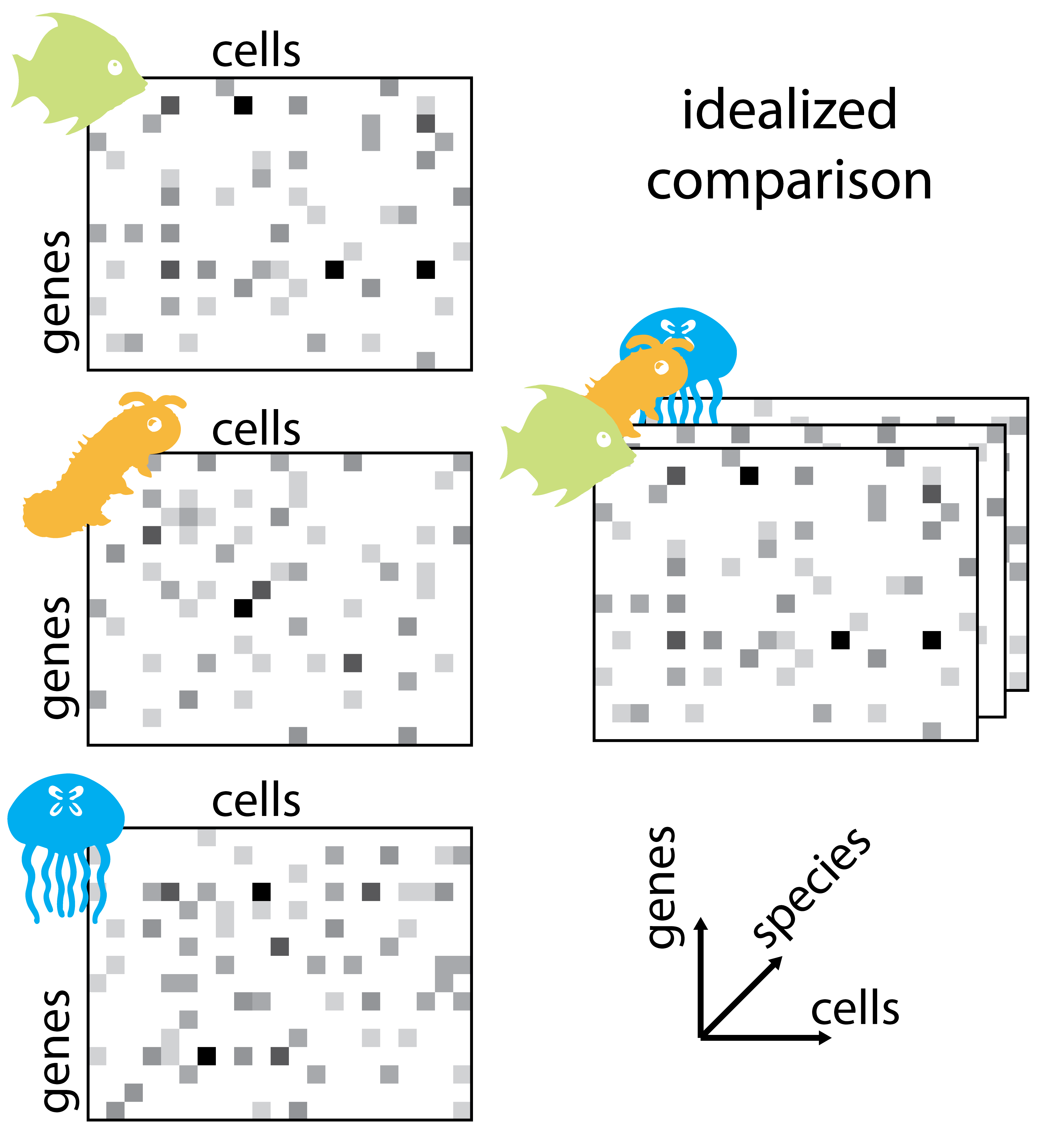}

}

\caption{Single cell experiments generate count matrices, shown here
with columns as cells and rows as genes. Higher expression counts for a
given gene in a given cell are depicted with darker shading. In an
idealized comparison, count matrices across species would be aligned to
form a three-dimensional tensor of expression across cells, genes, and
species. In reality we have no expectation of one-to-one correspondence
or independence for any the three axes. Instead, relationships between
species, genes, and cells are described by their respective evolutionary
histories, as depicted with phylogenies.}

\end{figure}

Questions about homology can be answered using
phylogenies\textsuperscript{12}. Species relationships are defined by
their shared ancestry, as depicted using a phylogeny of speciation
events (Figure 2). Gene homology is also defined by shared ancestry,
depicted using gene trees that contain nodes corresponding to either
speciation and gene duplication events. Cell homology inference requires
assessing the evolutionary relationships between cell
types\textsuperscript{12,13}, defined here as populations of cells
related via the process of cellular differentiation, and distinguishable
from one another e.g.~using molecular markers\textsuperscript{14}.
Relationships between cell types can be respresented with cell
phylogenies that, like gene trees, contain both speciation and
duplication nodes\textsuperscript{13}. As with genes, the evolutionary
relationships between cell types may be complex, as differentiation
trajectories drift, split, or are lost over evolutionary
time\textsuperscript{7,13,15}.

In this paper we consider each of these axes: across species, genes, and
cells. In each case we lay out challenges and solutions derived from a
phylogenetic comparative approach. We relate these solutions to methods
previously proposed in the literature, including
SAMap\textsuperscript{7} for pairwise alignment of cellular dimensional
reductions. Finally, we reconcile species trees, gene trees, cell
phylogenies and cell lineages as descriptions of the same concept--the
tree of cellular life.

\begin{figure}

{\centering \includegraphics[width=5.5in,height=\textheight]{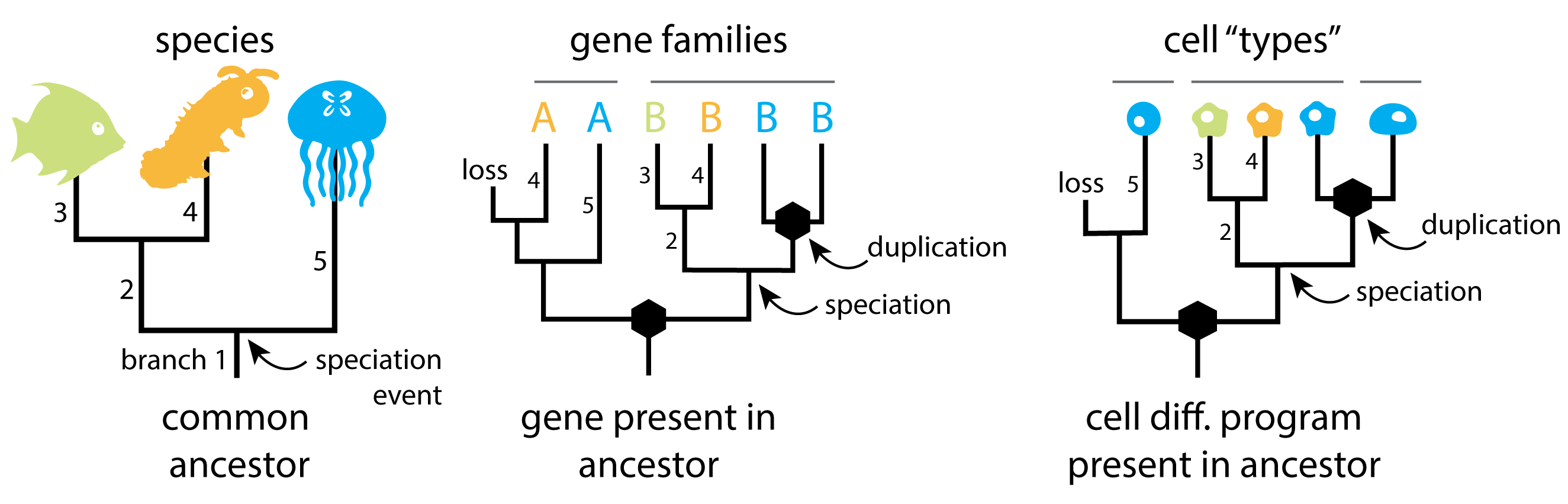}

}

\caption{Species phylogenies contain speciation events as nodes in a
bifurcating tree. Gene phylogenies contain both gene duplication events
(black hexagons) and speciation events (unmarked) at nodes. Cell
phylogenies also include both speciation and duplication events; here
duplication events represents a split in the program of cellular
development that leads to differentiated cell types\textsuperscript{13}.
Branches from the species phylogeny (numbered branches) can be found
within gene and cell phylogenies. Note that gene families are strictly
defined by ancestry, but cell types have historically been defined by
form, function, or patterns of gene expression\textsuperscript{15}. This
means groups of cells identified as the same ``type'' across species may
reflect paraphyletic groups\textsuperscript{11}, as depicted in the
second cell type in this tree.}

\end{figure}

\hypertarget{comparing-across-species}{%
\section{Comparing across species}\label{comparing-across-species}}

Shared ancestry between species will impact the results of all
cross-species analyses, and should therefore influence our expectations
and interpretations\textsuperscript{16}. For scRNA-seq data, this has
several implications: First, we expect species to be different from one
another, given that they have experienced evolutionary time since
diverging from their common ancestor. Therefore, by default we expect
there to be many differences in cellular gene expression across the
thousands of measurements in an scRNA-seq dataset. Second, we expect the
degree of difference to be correlated with time since the last common
ancestor. Our null expectation is that more closely related species will
have more similar cellular gene expression than more distantly related
species. The structure of this similarity can be approximated with a
species phylogeny calibrated to time.

Methods for the evolutionary comparison of scRNA-seq data have already
been proposed in packages like SAMap\textsuperscript{7}. These packages
have overcome significant challenges, such as how to account for
non-orthologous genes (see the description of gene comparisons below).
However, up to now they have relied on pairwise comparisons of species,
rather than phylogenetic relationships. The problems with pairwise
comparisons have been well-described elsewhere\textsuperscript{17}; in
summary, they result in pseudo-replication of evolutionary events. An
evolutionary comparative approach, on the other hand, maps evolutionary
changes to branches in the phylogeny\textsuperscript{8,10}. In this
approach:

\begin{itemize}
\tightlist
\item
  data are assigned to tips of a tree;
\item
  ancestral states are reconstructed using an evolutionary model;
\item
  evolutionary changes are calculated as differences between ancestral
  and descendant states;
\item
  the distribution of evolutionary changes along branches are analyzed
  and compared\textsuperscript{18}.
\end{itemize}

Shifting toward a phylogenetic approach to comparative scRNA-seq unlocks
new avenues of discovery, including tests of co-evolution of cellular
gene expression and other features of interest\textsuperscript{9}, as
well as evolutionary screens for signatures of correlated gene and cell
modules\textsuperscript{19}. In phylogenetic analyses, statistical power
depends on the number of independent evolutionary events rather the
absolute number of taxa\textsuperscript{8}. Therefore the choice of
which species to compare is critical, especially as we construct
comparisons to capture potential convergence.

One consideration when comparing species is the degree to which the
history of scientific study has favored certain organisms (e.g.~model
organisms)\textsuperscript{20}. This is especially relevant to
single-cell comparisons, as we have a much deeper knowledge about cell
and gene function for some species (e.g.~mice, humans) than others. This
creates a risk of bias toward observing described biological phenomena,
while missing the hidden biology in less well-studied
organisms\textsuperscript{20}. Consider the identification of ``novel''
cell types based on the absence of canonical marker genes. Because most
canonical marker genes were originally described in well-studied
species, cell type definitions that rely on these will be necessarily
less useful in the context of other species\textsuperscript{2}.

Technologies such as single-cell sequencing have great potential to even
the playing field by democratizing the types of data
collected\textsuperscript{2}. For example, scRNA-seq allows us to assay
all genes and thousands of cells, rather than a curated list of
candidates. To leverage this to full effect, we must acknowledge the
remaining filtering steps in our analyses, including how we identify
orthologous gene sequences and how we label cell types.

\hypertarget{comparing-across-genes}{%
\section{Comparing across genes}\label{comparing-across-genes}}

Due to gene duplication and loss, there is usually not a one-to-one
correspondence between genes across species\textsuperscript{21}.
Instead, evolutionary histories of genes are depicted using gene trees
(Figure 2). Pairs of tips in gene trees may be labeled as ``orthologs''
or ``paralogs'' based on whether they descend from a node corresponding
to a speciation or gene duplication event\textsuperscript{22}. Gene
duplication happens both at the individual gene level, or in bulk via
whole or partial genome duplication\textsuperscript{21}. Gene loss means
that comparative scRNA-seq matrices may be sparse not only due to
failure to detect, but also because genes in one species don't exist in
another.

Many cross-species comparisons have confronted the challenge of finding
equivalent genes across species\textsuperscript{23}. These often start
by restricting analyses to sets of one-to-one
orthologs\textsuperscript{24}. There are several problems with this
approach\textsuperscript{22}: {[}1{]} One-to-one orthologs are only
well-described for a small set of very well annotated
genomes\textsuperscript{23}. {[}2{]} The number of one-to-one orthologs
decreases rapidly as we both add species to our comparison, and as we
compare across deeper evolutionary distances\textsuperscript{7}. {[}3{]}
The subset of genes that can be described by one-to-one orthologs is not
randomly drawn from across the genome, they are enriched for
indispensable genes under single-copy control\textsuperscript{25}.

New tools like SAMap\textsuperscript{7} expand the analytical approach
beyond one-to-one orthologs to the set of all homologs across
species\textsuperscript{7}. Homolog groups are identified with a
clustering algorithm by which gene separated into groups with strong
sequence or expression similarity. These may include more than one
representative gene per species. Gene trees can be inferred for these
gene families, and duplication events can be mapped to individual nodes
in the gene tree.

How can we compare cellular expression measures across groups of
homologous genes? One option is to use summary statistics, such as the
sum or average expression per species for genes within a homology
group\textsuperscript{26}. However these statistics might obscure or
average over real biological variation in expression that arose
subsequent to a duplication event (among paralogs)\textsuperscript{19}.
An alternative is to connect genes via a similarity matrix, and then
make all-by-all comparisons that are weighted based on putative
homology, as in the approach in SAMap\textsuperscript{7}.

A third approach is to reconstruct changes in cellular expression along
gene trees, rather than the species tree\textsuperscript{10,27}. Here
evolutionary changes are associated with branches descending from either
speciation or duplication events. Such an approach has been demonstrated
by Munro \emph{et. al.} for bulk RNA-sequencing\textsuperscript{27}, as
follows:

\begin{itemize}
\tightlist
\item
  cellular expression data are assigned to tips of a gene tree;
\item
  ancestral states and evolutionary changes are calculated, as above;
\item
  equivalent branches between trees are identified using ``species
  branch filtering''\textsuperscript{27}. Branches between speciation
  events can be unambiguously equated across trees based on the
  composition of their descendant tips (see numbered branches in Figure
  2);
\item
  changes across equivalent branches of a cell tree are analyzed
  (e.g.~to identify significant changes, signatures of correlation,
  etc).
\end{itemize}

Mapping cellular gene expression data to branches of a gene tree
sidesteps the problem of finding sets of orthologs by incorporating the
history of gene duplication and loss into the analytical framework.

\hypertarget{comparing-across-cells}{%
\section{Comparing across cells}\label{comparing-across-cells}}

Like with genes, there is usually not an expectation of a one-to-one
correspondence between cells across species. We can rarely equate
individual cells across species, with notable exceptions such as the
zygote, or the cells of certain eutelic species, with an invariant
number of cellular divisions. Instead we typically consider the homology
of groups of cells (cell types), with the hypothesis that the cell
developmental programs that give rise to these groups are derived from a
program present in the shared ancestor\textsuperscript{11}.

We also don't expect a one-to-one correspondence between cell types
across species. As with genes, cell types may be gained or lost over
evolutionary time. The relationships between cell types across species
can been described using phylogenetic trees. These cell phylogenies are
distinct from cell lineages--the bifurcating trees that describe
cellular divisions within an individual developmental history. Nodes in
cell lineages represent cell divisions, while nodes in cell phylogenies
represent either speciation events or splits in differentiation programs
leading to novel cell types (Figure 2).

The term cell ``type'' has been used for several distinct
concepts\textsuperscript{15}, including cells defined and distinguished
by their position in a tissue, their form, function, or in the case of
scRNA-seq, their relative expression profiles that fall into distinct
clusters\textsuperscript{14}. Homology of structures across species is
often inferred using many of the same criteria: position, form,
function, and gene expression patterns\textsuperscript{28}. The fact
that the same principles are used for inferring cell types and cell
homologies presents both an opportunity and an obstacle for comparative
scRNA-seq. We can potentially leverage the same methods we use for
identifying clusters of cells within species to identify clusters of
cells across species. This could be done simultaneously, inferring a
joint cell atlas in a shared expression space\textsuperscript{7}, or it
could be done individually for each species and subsequently
merged\textsuperscript{2,23}. In either case, this inference requires
contending with the complex evolutionary histories between genes and
species described above.

One obstacle is that, because cell types are not typically defined
according to evolutionary relationships\textsuperscript{15}, cells
labeled as the same type across species may constitute paraphyletic
groups\textsuperscript{11}. A solution is to use methods for
reconstructing evolutionary relationships to infer the cell
tree\textsuperscript{15,29} (Figure 2). This method is robust to complex
evolutionary histories (e.g.~duplication and loss), and has the
additional advantage of generating a tree, comparable to a species or
gene tree, onto which cellular characters can be mapped and their
evolution described\textsuperscript{15}. As described in Mah and Dunn,
2023\textsuperscript{30}, tools for tree building based on character
data can be applied to the high-dimensional data of gene expression to
successfully reconstruct a cell tree. In this approach:

\begin{itemize}
\tightlist
\item
  cell trees are inferred;
\item
  gene expression data are assigned to tips of the cell tree;
\item
  ancestral states and evolutionary changes are calculated, as above;
\item
  changes along branches are analyzed (e.g.~to identify changes in gene
  expression associated with the evolution of novel cell types)
\end{itemize}

Another obstacle is that there are reported batch
effects\textsuperscript{26} across single-cell experiments which may
need to be accounted for via integration\textsuperscript{23}. However
our null expectation is that species are different from one another.
Naive batch integration practices have no method for distinguishing
technical effects from the real biological differences that are the
target of study in comparative scRNA-seq\textsuperscript{23}. Other
approaches (e.g.~\texttt{LIGER}\textsuperscript{31},
\texttt{Seurat}\textsuperscript{32}) have been reported as able to
distinguish and characterize species-specific
differences\textsuperscript{23}. Given that we are still developing null
hypotheses\textsuperscript{16} for how much variation in expression we
expect to observe across species\textsuperscript{19}, we hold that
cross-species integration should be treated with caution until
elucidation of the approach can robustly target and strictly remove
technical batch effects.

A final obstacle is that cell identities and homologies may be more
complex than can be accurately captured by categorizations into discrete
clusters or ``types''\textsuperscript{14,15}. Single-cell experiments
that include both progenitor and differentiated cells can highlight the
limits of clustering algorithms\textsuperscript{33}. In experiments that
capture cells along a differentiation trajectory, there may or may not
be obvious boundaries for distinguishing cell populations. In cases
where boundaries are arbitrary, the number of clusters, and therefore
the abundance of cells within a cluster will depend on technical and not
biological inputs, such as the resolution parameter that the user
predetermines for the clustering algorithm.

A solution is to define homology for the entire differentiation
trajectory, rather than individual clusters of
cells\textsuperscript{26}. This may be accomplished by defining anchor
points where trajectories overlap in the expression of homologous genes,
while allowing for trajectories to have drifted or split over
evolutionary time, such that sections of the trajectories no longer
overlap\textsuperscript{15}. Cellular homologies within a trajectory may
be more difficult to infer, as this requires contending with potential
heterochronic changes to differentiation (e.g.~as cell differentiation
evolves, genes may become expressed relatively earlier or later in the
process)\textsuperscript{26}.

\hypertarget{scrna-seq-data}{%
\section{scRNA-seq data}\label{scrna-seq-data}}

Single-cell comparisons potentially draw on a broad range of
phylogenetic comparative methods for different data types, including
binary, discrete, continuous, and categorical data\textsuperscript{34}
(Figure 3). The primary data structure of scRNA-seq is a matrix of
integers, representing counts of transcripts or unique molecular
identifiers for a given gene within a given cell\textsuperscript{35}. In
a typical scRNA-seq analysis, this count matrix is passed through a
pipeline of normalization, transformation, dimensional reduction, and
clustering\textsuperscript{36,37}. The decisions of when during this
pipeline to draw a comparison determines data type, questions we can
address, and caveats we must consider.

\begin{figure}

{\centering \includegraphics[width=5.5in,height=\textheight]{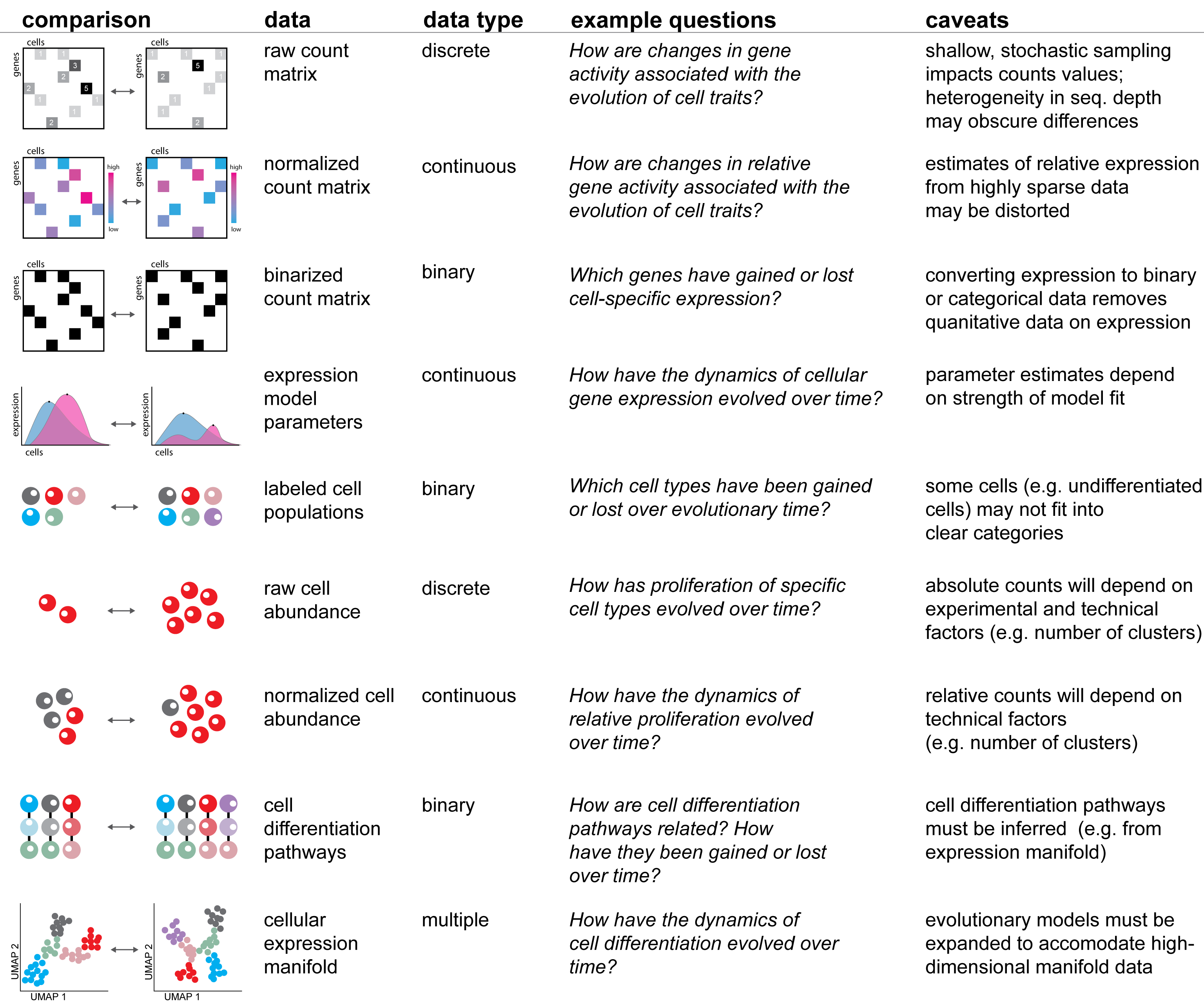}

}

\caption{Types of scRNA-seq data that can be mapped onto a tree, with
example questions that can be addressed and caveats to be considered.}

\end{figure}

\hypertarget{gene-expression}{%
\subsection{Gene expression}\label{gene-expression}}

Unlike bulk RNA-sequencing, where counts are typically distributed
across a few to dozens of samples, scRNA-seq counts are distributed
across thousands of cells. The result is that scRNA-seq count matrices
are often shallow and sparse\textsuperscript{38}. In our recent
paper\textsuperscript{35}, we highlighted that the vast majority of UMI
counts in standard scRNA-seq datasets (often \textgreater95\%) are
either 0, 1, or 2. These values are discrete, low-integer numbers, and
not continuous measurements. The high-dimensionality and sparse nature
of single-cell data present a unique challenge when considering
cross-species comparisons\textsuperscript{2}.

In a standard scRNA-seq approach, expression values are analyzed
following depth normalization and other transformations. With depth
normalization, counts are converted from discrete, absolute measures to
continuous, relative ones, though our instruments don't actually
quantify relative expression. There is a growing concern that this and
other transformations are inadequate for the sparse and shallow
sequencing data introduced here\textsuperscript{39,40}. Further
transformations of the data, such as log-transformation or variance
rescaling, introduce additional distortions that may obscure real
biological differences between species.

Alternatively, counts can be compared across species directly, without
normalization or transformation\textsuperscript{35}. There are two
potential drawbacks to this approach: First, count values are influenced
by stochasticity, due to the shallow nature of sequencing, resulting in
uncertainty around integer values. Second, cells are not sequenced to a
standard depth. Comparing raw counts does not take this heterogeneity
into account, though our recently published approach\textsuperscript{35}
describes how this might be accomplished using a restricted algebra to
analyze counts.

An alternative is to transform count values to a binary or categorical
trait\textsuperscript{41}, for example, binning counts into ``on'' and
``off'' based on a threshold value, and then model the evolution of
these states on a tree. Analyzing expression as a binary or categorical
trait eliminates some of the quantitative power of scRNA-seq, but still
allows us to address interesting questions about the evolution of
expression dynamics within and across cell types.

\hypertarget{models-of-expression}{%
\subsection{Models of expression}\label{models-of-expression}}

A promising avenue for scRNA-seq is using generalized linear models to
analyze expression\textsuperscript{40,42,43}. These models describe
expression as a continuous trait and incorporate the sampling process
using a Poisson or other distribution, avoiding normalization and
transformation, and returning fitted estimates of relative expression.
These estimates can be compared using models that describe continuous
trait evolution. One feature of generalized linear models models is they
can report uncertainty values for our estimates of relative expression,
which can then be passed along to phylogenetic methods to assess
confidence in the evolutionary conclusions drawn.

\hypertarget{cell-diversity}{%
\subsection{Cell diversity}\label{cell-diversity}}

In a standard scRNA-seq approach cells are analyzed in a reduced
dimensional space and clustered by patterns of gene
expression\textsuperscript{37}. There are several types of cellular data
that can be compared. The evolution of the presence or absence of cell
types can be modeled as a binary trait. When cell type labels are
unambiguously assigned, this approach can answer questions about when
cell types evolved and are lost. Such a comparison is hampered, however,
when cells do not fall into discrete categories\textsuperscript{14}, or
when equivalent cell types cannot be identified across species due to
substantial divergence in gene expression patterns. An alternative is to
model the evolution of cell differentiation pathways as a binary trait
on a tree, to ask when pathways, rather than types, evolved and have
been lost.

Similarly, the abundance of cells of a given type might be compared
across species, for example to ask how how dynamics of cell
proliferation have evolved. The number of cells within a cluster,
however, can be influenced by technical features of the experiment, such
as the total number of clusters identified (often influenced by
user-supplied parameters), as well as where cluster boundaries are
defined. An alternative is to compare relative cell abundance values,
which may account for experimental factors but will still be susceptible
to variation in how clusters are determined.

\hypertarget{cellular-manifolds}{%
\subsection{Cellular manifolds}\label{cellular-manifolds}}

One area for further development are methods that can model the
evolution of the entire manifold of cellular gene expression on an
evolutionary tree. Practically, this might be accomplished by
parameterizing the manifold, for example by calculating measures of
manifold shape and structure such as distances between cells in a
reduced dimensional space. The evolution of such parameters could be
studied by analyzing them as characters on a phylogenetic tree.

Alternatively, we can envision a method in which we reconstruct entire
ancestral landscapes of cellular gene expression, and then describe how
this landscape has been reshaped over evolutionary time. Such an
approach would require an expansion of existing phylogenetic comparative
models to ones that can incorporate many thousands of dimensions. It
would also likely require dense taxonomic sampling to build robust
reconstructions.

\hypertarget{discussion}{%
\section{Discussion}\label{discussion}}

Comparative single-cell RNA sequencing spans the fields of evolutionary,
developmental, and cellular biology. Phylogenetic trees--branching
structures depicting relationships across time--are the common
denominator of these fields. Taking a step back reveals that many of the
trees we typically encounter, such as species phylogenies, gene
phylogenies, cell phylogenies, and cell fate maps, can be reconciled as
part of a larger whole (Figure 4). Because all cellular life is related
via an unbroken chain of cellular divisions, species phylogenies and
cell fate maps are two representations of the same larger phenomenon,
visualized at vastly different scales. Gene trees and cell trees
(i.e.~cell phylogenies) depict the evolution of specific characters
(genes and cells) across populations within a species tree. These
characters may have discordant evolutionary histories due to patterns of
gene and cell duplication, loss, and incomplete sorting across
populations.

\begin{figure}

{\centering \includegraphics{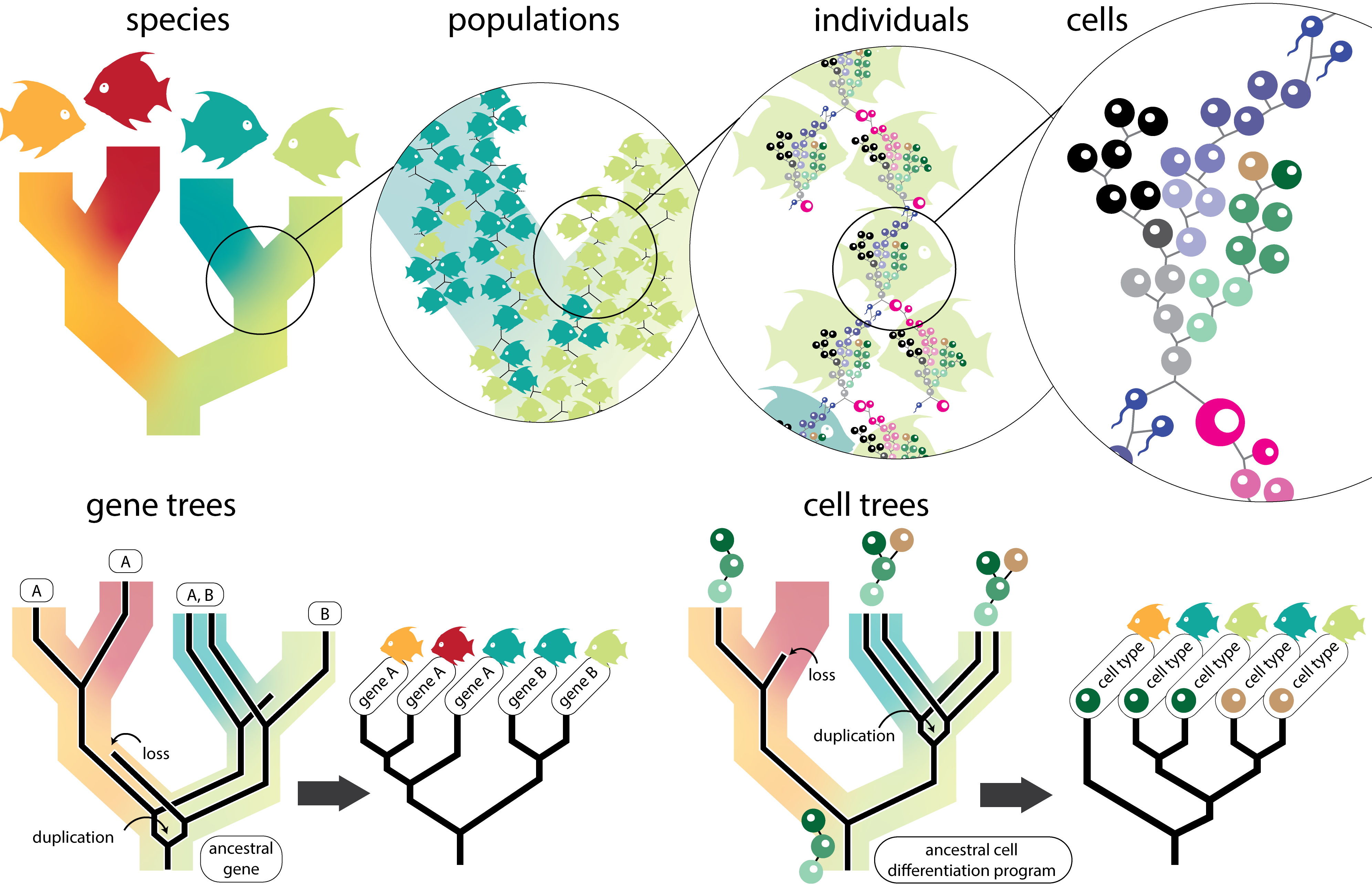}

}

\caption{All cellular life is related by an unbroken chain of cell
divisions. Species phylogenies describe the relationship between
populations. Populations are themselves a description of the
genealogical relationships between individuals. Peering even closer
reveals that each individual consists of a lineage of cells, connected
to other individuals via reproductive cells. Therefore, species trees,
genealogies, and cell lineages are all descriptions of the same
concept--the tree of life--but at different scales. Gene trees and cell
trees (i.e.~cellular phylogenies) describe the evolutionary histories of
specific characters within the tree of life. These trees may be
discordant with species trees due to duplication, loss, and incomplete
sorting in populations.}

\end{figure}

The synthesis of species, gene, and cell trees makes several key points
clear. First, phylogenetic trees are essential for testing hypotheses
about cellular gene expression evolution. Mapping single-cell data to
trees, whether gene trees, cell trees, or species trees, allows us to
build statistical tests of co-evolution, diversification, and
convergence. The choice of which trees to use for mapping data will be
determined by the questions we hope to answer. For example, mapping
cellular expression data to gene trees would allow us to test whether
expression evolves differently following gene duplication events
(i.e.~to test the ortholog conjecture\textsuperscript{44}).

Second, because the fields of evolutionary, developmental, and cellular
biology study the same phenomena at different scales, there is a
potential benefit from sharing methods. In the case of single-cell
sequencing, building evolutionary context around data can prove
essential for understanding the fundamental biology, including how to
interpret cell types and cellular differentiation trajectories, and how
to reconcile gene relationships. An evolutionary perspective is also
critical for building robust null expectations of how much variation we
might expect to observe across species\textsuperscript{16}, which will
allow us to interpret the significance of results as new species atlases
come to light.

Methods that infer and incorporate trees are essential not only for
evolutionary biology, but for developmental and cellular biology as
well. As single-cell data become increasingly available, rather than
reinvent methods for building cell trees or comparing across cellular
network diagrams, we can draw approaches from the extensive and robust
fields of phylogenetic inference and phylogenetic comparative methods.
These approaches include Bayesian and Maximum Likelihood inference of
trees, evolutionary models, ancestral state reconstruction, character
state matrices, phylogenetic hypothesis testing, among many
others\textsuperscript{45--47}.

Biology has benefited in the past from syntheses of disparate fields of
study, including the modern synthesis of Darwinian evolution and
Mendelian genetics\textsuperscript{48}, and the synthesis of evolution
and development in the field of evo-devo\textsuperscript{49}. With the
advent and commercialization of technologies like single-cell
sequencing, there is broadened opportunity for new
syntheses\textsuperscript{50}. Rich and complex datasets are
increasingly available from understudied branches on the tree of life,
and comparisons between species will invariably invoke evolutionary
questions. By integrating phylogenetic thinking across fields, we can
start to answer these questions and raise new ones.

\hypertarget{acknowledgments}{%
\section{Acknowledgments}\label{acknowledgments}}

We thank Daniel Stadtmauer, Namrata Ahuja, Seth Donoughe and other
members of the Dunn lab for helpful conversation and comments on an
initial version of the manuscript.

\hypertarget{declaration-of-interests}{%
\section{Declaration of interests}\label{declaration-of-interests}}

The authors declare no competing interests.

\hypertarget{references}{%
\section*{References}\label{references}}
\addcontentsline{toc}{section}{References}

\hypertarget{refs}{}
\begin{CSLReferences}{0}{0}
\leavevmode\vadjust pre{\hypertarget{ref-gawad2016single}{}}%
\CSLLeftMargin{1. }%
\CSLRightInline{Gawad, C., Koh, W. \& Quake, S. R. Single-cell genome
sequencing: Current state of the science. \emph{Nature Reviews Genetics}
\textbf{17}, 175--188 (2016).}

\leavevmode\vadjust pre{\hypertarget{ref-tanay2021evolutionary}{}}%
\CSLLeftMargin{2. }%
\CSLRightInline{Tanay, A. \& Sebé-Pedrós, A. Evolutionary cell type
mapping with single-cell genomics. \emph{Trends in Genetics}
\textbf{37}, 919--932 (2021).}

\leavevmode\vadjust pre{\hypertarget{ref-sebe2018early}{}}%
\CSLLeftMargin{3. }%
\CSLRightInline{Sebé-Pedrós, A. \emph{et al.} Early metazoan cell type
diversity and the evolution of multicellular gene regulation.
\emph{Nature Ecology \& Evolution} \textbf{2}, 1176--1188 (2018).}

\leavevmode\vadjust pre{\hypertarget{ref-li2021single}{}}%
\CSLLeftMargin{4. }%
\CSLRightInline{Li, P. \emph{et al.} Single-cell analysis of schistosoma
mansoni identifies a conserved genetic program controlling germline stem
cell fate. \emph{Nature Communications} \textbf{12}, 485 (2021).}

\leavevmode\vadjust pre{\hypertarget{ref-levy2021stony}{}}%
\CSLLeftMargin{5. }%
\CSLRightInline{Levy, S. \emph{et al.} A stony coral cell atlas
illuminates the molecular and cellular basis of coral symbiosis,
calcification, and immunity. \emph{Cell} \textbf{184}, 2973--2987
(2021).}

\leavevmode\vadjust pre{\hypertarget{ref-hulett2022acoel}{}}%
\CSLLeftMargin{6. }%
\CSLRightInline{Hulett, R. E. \emph{et al.} Acoel single-cell atlas
reveals expression dynamics and heterogeneity of a pluripotent stem cell
population. \emph{BioRxiv} 2022--02 (2022).}

\leavevmode\vadjust pre{\hypertarget{ref-tarashansky2021mapping}{}}%
\CSLLeftMargin{7. }%
\CSLRightInline{Tarashansky, A. J. \emph{et al.} Mapping single-cell
atlases throughout metazoa unravels cell type evolution. \emph{Elife}
\textbf{10}, e66747 (2021).}

\leavevmode\vadjust pre{\hypertarget{ref-smith2020phylogenetics}{}}%
\CSLLeftMargin{8. }%
\CSLRightInline{Smith, S. D., Pennell, M. W., Dunn, C. W. \& Edwards, S.
V. Phylogenetics is the new genetics (for most of biodiversity).
\emph{Trends in Ecology \& Evolution} \textbf{35}, 415--425 (2020).}

\leavevmode\vadjust pre{\hypertarget{ref-romero2012comparative}{}}%
\CSLLeftMargin{9. }%
\CSLRightInline{Romero, I. G., Ruvinsky, I. \& Gilad, Y. Comparative
studies of gene expression and the evolution of gene regulation.
\emph{Nature Reviews Genetics} \textbf{13}, 505--516 (2012).}

\leavevmode\vadjust pre{\hypertarget{ref-dunn2013phylogenetic}{}}%
\CSLLeftMargin{10. }%
\CSLRightInline{Dunn, C. W., Luo, X. \& Wu, Z. Phylogenetic analysis of
gene expression. \emph{Integrative and Comparative Biology} \textbf{53},
847--856 (2013).}

\leavevmode\vadjust pre{\hypertarget{ref-arendt2016origin}{}}%
\CSLLeftMargin{11. }%
\CSLRightInline{Arendt, D. \emph{et al.} The origin and evolution of
cell types. \emph{Nature Reviews Genetics} \textbf{17}, 744--757
(2016).}

\leavevmode\vadjust pre{\hypertarget{ref-wagner2014homology}{}}%
\CSLLeftMargin{12. }%
\CSLRightInline{Wagner, G. P. \emph{Homology, genes, and evolutionary
innovation}. (Princeton University Press, 2014).}

\leavevmode\vadjust pre{\hypertarget{ref-arendt2008evolution}{}}%
\CSLLeftMargin{13. }%
\CSLRightInline{Arendt, D. The evolution of cell types in animals:
Emerging principles from molecular studies. \emph{Nature Reviews
Genetics} \textbf{9}, 868--882 (2008).}

\leavevmode\vadjust pre{\hypertarget{ref-domcke2023reference}{}}%
\CSLLeftMargin{14. }%
\CSLRightInline{Domcke, S. \& Shendure, J. A reference cell tree will
serve science better than a reference cell atlas. \emph{Cell}
\textbf{186}, 1103--1114 (2023).}

\leavevmode\vadjust pre{\hypertarget{ref-kin2015inferring}{}}%
\CSLLeftMargin{15. }%
\CSLRightInline{Kin, K. Inferring cell type innovations by phylogenetic
methods---concepts, methods, and limitations. \emph{Journal of
Experimental Zoology Part B: Molecular and Developmental Evolution}
\textbf{324}, 653--661 (2015).}

\leavevmode\vadjust pre{\hypertarget{ref-church2020null}{}}%
\CSLLeftMargin{16. }%
\CSLRightInline{Church, S. H. \& Extavour, C. G. Null hypotheses for
developmental evolution. \emph{Development} \textbf{147}, dev178004
(2020).}

\leavevmode\vadjust pre{\hypertarget{ref-dunn2018pairwise}{}}%
\CSLLeftMargin{17. }%
\CSLRightInline{Dunn, C. W., Zapata, F., Munro, C., Siebert, S. \&
Hejnol, A. Pairwise comparisons across species are problematic when
analyzing functional genomic data. \emph{Proceedings of the National
Academy of Sciences} \textbf{115}, E409--E417 (2018).}

\leavevmode\vadjust pre{\hypertarget{ref-felsenstein1985phylogenies}{}}%
\CSLLeftMargin{18. }%
\CSLRightInline{Felsenstein, J. Phylogenies and the comparative method.
\emph{The American Naturalist} \textbf{125}, 1--15 (1985).}

\leavevmode\vadjust pre{\hypertarget{ref-church2023evolution}{}}%
\CSLLeftMargin{19. }%
\CSLRightInline{Church, S. H., Munro, C., Dunn, C. W. \& Extavour, C. G.
The evolution of ovary-biased gene expression in hawaiian drosophila.
\emph{PLoS Genetics} \textbf{19}, e1010607 (2023).}

\leavevmode\vadjust pre{\hypertarget{ref-dunn2015hidden}{}}%
\CSLLeftMargin{20. }%
\CSLRightInline{Dunn, C. W., Leys, S. P. \& Haddock, S. H. The hidden
biology of sponges and ctenophores. \emph{Trends in Ecology \&
Evolution} \textbf{30}, 282--291 (2015).}

\leavevmode\vadjust pre{\hypertarget{ref-sankoff2001gene}{}}%
\CSLLeftMargin{21. }%
\CSLRightInline{Sankoff, D. Gene and genome duplication. \emph{Current
Opinion in Genetics \& Development} \textbf{11}, 681--684 (2001).}

\leavevmode\vadjust pre{\hypertarget{ref-dunn2016comparative}{}}%
\CSLLeftMargin{22. }%
\CSLRightInline{Dunn, C. W. \& Munro, C. Comparative genomics and the
diversity of life. \emph{Zoologica Scripta} \textbf{45}, 5--13 (2016).}

\leavevmode\vadjust pre{\hypertarget{ref-shafer2019cross}{}}%
\CSLLeftMargin{23. }%
\CSLRightInline{Shafer, M. E. Cross-species analysis of single-cell
transcriptomic data. \emph{Frontiers in Cell and Developmental Biology}
\textbf{7}, 175 (2019).}

\leavevmode\vadjust pre{\hypertarget{ref-stuart2019integrative}{}}%
\CSLLeftMargin{24. }%
\CSLRightInline{Stuart, T. \& Satija, R. Integrative single-cell
analysis. \emph{Nature Reviews Genetics} \textbf{20}, 257--272 (2019).}

\leavevmode\vadjust pre{\hypertarget{ref-waterhouse2011correlating}{}}%
\CSLLeftMargin{25. }%
\CSLRightInline{Waterhouse, R. M., Zdobnov, E. M. \& Kriventseva, E. V.
Correlating traits of gene retention, sequence divergence, duplicability
and essentiality in vertebrates, arthropods, and fungi. \emph{Genome
Biology and Evolution} \textbf{3}, 75--86 (2011).}

\leavevmode\vadjust pre{\hypertarget{ref-marioni2017single}{}}%
\CSLLeftMargin{26. }%
\CSLRightInline{Marioni, J. C. \& Arendt, D. How single-cell genomics is
changing evolutionary and developmental biology. \emph{Annual Review of
Cell and Developmental Biology} \textbf{33}, 537--553 (2017).}

\leavevmode\vadjust pre{\hypertarget{ref-munro2022evolution}{}}%
\CSLLeftMargin{27. }%
\CSLRightInline{Munro, C., Zapata, F., Howison, M., Siebert, S. \& Dunn,
C. W. Evolution of gene expression across species and specialized zooids
in siphonophora. \emph{Molecular Biology and Evolution} \textbf{39},
msac027 (2022).}

\leavevmode\vadjust pre{\hypertarget{ref-wagner2000developmental}{}}%
\CSLLeftMargin{28. }%
\CSLRightInline{Wagner, G. P., Chiu, C.-H. \& Laubichler, M.
Developmental evolution as a mechanistic science: The inference from
developmental mechanisms to evolutionary processes. \emph{American
Zoologist} \textbf{40}, 819--831 (2000).}

\leavevmode\vadjust pre{\hypertarget{ref-wang2021tracing}{}}%
\CSLLeftMargin{29. }%
\CSLRightInline{Wang, J. \emph{et al.} Tracing cell-type evolution by
cross-species comparison of cell atlases. \emph{Cell Reports}
\textbf{34}, 108803 (2021).}

\leavevmode\vadjust pre{\hypertarget{ref-mah2023reconstructing}{}}%
\CSLLeftMargin{30. }%
\CSLRightInline{Mah, J. L. \& Dunn, C. Reconstructing cell type
evolution across species through cell phylogenies of single-cell RNAseq
data. \emph{bioRxiv} 2023--05 (2023).}

\leavevmode\vadjust pre{\hypertarget{ref-welch2019single}{}}%
\CSLLeftMargin{31. }%
\CSLRightInline{Welch, J. D. \emph{et al.} Single-cell multi-omic
integration compares and contrasts features of brain cell identity.
\emph{Cell} \textbf{177}, 1873--1887 (2019).}

\leavevmode\vadjust pre{\hypertarget{ref-butler2018integrating}{}}%
\CSLLeftMargin{32. }%
\CSLRightInline{Butler, A., Hoffman, P., Smibert, P., Papalexi, E. \&
Satija, R. Integrating single-cell transcriptomic data across different
conditions, technologies, and species. \emph{Nature Biotechnology}
\textbf{36}, 411--420 (2018).}

\leavevmode\vadjust pre{\hypertarget{ref-tritschler2019concepts}{}}%
\CSLLeftMargin{33. }%
\CSLRightInline{Tritschler, S. \emph{et al.} Concepts and limitations
for learning developmental trajectories from single cell genomics.
\emph{Development} \textbf{146}, dev170506 (2019).}

\leavevmode\vadjust pre{\hypertarget{ref-cornwell2017phylogenetic}{}}%
\CSLLeftMargin{34. }%
\CSLRightInline{Cornwell, W. \& Nakagawa, S. Phylogenetic comparative
methods. \emph{Current Biology} \textbf{27}, R333--R336 (2017).}

\leavevmode\vadjust pre{\hypertarget{ref-church2022normalizing}{}}%
\CSLLeftMargin{35. }%
\CSLRightInline{Church, S. H., Mah, J. L., Wagner, G. \& Dunn, C.
Normalizing need not be the norm: Count-based math for analyzing
single-cell data. \emph{bioRxiv} 2022--06 (2022).}

\leavevmode\vadjust pre{\hypertarget{ref-satija2015spatial}{}}%
\CSLLeftMargin{36. }%
\CSLRightInline{Satija, R., Farrell, J. A., Gennert, D., Schier, A. F.
\& Regev, A. Spatial reconstruction of single-cell gene expression data.
\emph{Nature Biotechnology} \textbf{33}, 495--502 (2015).}

\leavevmode\vadjust pre{\hypertarget{ref-luecken2019current}{}}%
\CSLLeftMargin{37. }%
\CSLRightInline{Luecken, M. D. \& Theis, F. J. Current best practices in
single-cell RNA-seq analysis: A tutorial. \emph{Molecular Systems
Biology} \textbf{15}, e8746 (2019).}

\leavevmode\vadjust pre{\hypertarget{ref-liu2016single}{}}%
\CSLLeftMargin{38. }%
\CSLRightInline{Liu, S. \& Trapnell, C. Single-cell transcriptome
sequencing: Recent advances and remaining challenges.
\emph{F1000Research} \textbf{5}, (2016).}

\leavevmode\vadjust pre{\hypertarget{ref-hicks2018missing}{}}%
\CSLLeftMargin{39. }%
\CSLRightInline{Hicks, S. C., Townes, F. W., Teng, M. \& Irizarry, R. A.
Missing data and technical variability in single-cell
{R}{N}{A}-sequencing experiments. \emph{Biostatistics} \textbf{19},
562--578 (2018).}

\leavevmode\vadjust pre{\hypertarget{ref-townes2019feature}{}}%
\CSLLeftMargin{40. }%
\CSLRightInline{Townes, F. W., Hicks, S. C., Aryee, M. J. \& Irizarry,
R. A. Feature selection and dimension reduction for single-cell RNA-seq
based on a multinomial model. \emph{Genome Biology} \textbf{20}, 1--16
(2019).}

\leavevmode\vadjust pre{\hypertarget{ref-qiu2020embracing}{}}%
\CSLLeftMargin{41. }%
\CSLRightInline{Qiu, P. Embracing the dropouts in single-cell RNA-seq
analysis. \emph{Nature Communications} \textbf{11}, 1169 (2020).}

\leavevmode\vadjust pre{\hypertarget{ref-hafemeister2019normalization}{}}%
\CSLLeftMargin{42. }%
\CSLRightInline{Hafemeister, C. \& Satija, R. Normalization and variance
stabilization of single-cell RNA-seq data using regularized negative
binomial regression. \emph{Genome Biology} \textbf{20}, 296 (2019).}

\leavevmode\vadjust pre{\hypertarget{ref-ahlmann2020glmgampoi}{}}%
\CSLLeftMargin{43. }%
\CSLRightInline{Ahlmann-Eltze, C. \& Huber, W. glmGamPoi: Fitting
gamma-poisson generalized linear models on single cell count data.
\emph{Bioinformatics} \textbf{36}, 5701--5702 (2020).}

\leavevmode\vadjust pre{\hypertarget{ref-nehrt2011testing}{}}%
\CSLLeftMargin{44. }%
\CSLRightInline{Nehrt, N. L., Clark, W. T., Radivojac, P. \& Hahn, M. W.
Testing the ortholog conjecture with comparative functional genomic data
from mammals. \emph{PLoS Computational Biology} \textbf{7}, e1002073
(2011).}

\leavevmode\vadjust pre{\hypertarget{ref-swofford19961996}{}}%
\CSLLeftMargin{45. }%
\CSLRightInline{Swofford, D., Olsen, G., Waddell, P. \& Hillis, D.
Phylogenetic inference. in \emph{Molecular systematics} 407--514
(Sinauer, 1996).}

\leavevmode\vadjust pre{\hypertarget{ref-baum2008phylogenics}{}}%
\CSLLeftMargin{46. }%
\CSLRightInline{Baum, D. A. \& Offner, S. Phylogenics \& tree-thinking.
\emph{The American Biology Teacher} \textbf{70}, 222--229 (2008).}

\leavevmode\vadjust pre{\hypertarget{ref-harmon2019phylogenetic}{}}%
\CSLLeftMargin{47. }%
\CSLRightInline{Harmon, L. \emph{Phylogenetic comparative methods}.
(Independent, 2019).}

\leavevmode\vadjust pre{\hypertarget{ref-huxley1942evolution}{}}%
\CSLLeftMargin{48. }%
\CSLRightInline{Huxley, J. \emph{Evolution. The modern synthesis.}
(George Alien \& Unwin Ltd., 1942).}

\leavevmode\vadjust pre{\hypertarget{ref-carroll2008evo}{}}%
\CSLLeftMargin{49. }%
\CSLRightInline{Carroll, S. B. Evo-devo and an expanding evolutionary
synthesis: A genetic theory of morphological evolution. \emph{Cell}
\textbf{134}, 25--36 (2008).}

\leavevmode\vadjust pre{\hypertarget{ref-abouheif2014eco}{}}%
\CSLLeftMargin{50. }%
\CSLRightInline{Abouheif, E. \emph{et al.} Eco-evo-devo: The time has
come. \emph{Ecological Genomics: Ecology and the evolution of genes and
genomes} \textbf{781}, 107--125 (2014).}

\end{CSLReferences}

\end{document}